\documentclass[aps,pra,preprint,groupedaddress,showpacs]{revtex4}

\usepackage{graphicx}
\usepackage{amsmath}
\usepackage{amssymb}
\usepackage{epsfig}

\begin{document}
\newcommand{\ped}[1]{_{\text{#1}}}
\newcommand{\diff}[1]{\text{d}#1}
%\preprint{APS/123-QED}

\title{Controlled Shock Shells and Intracluster Fusion Reactions in the Explosion of Large Clusters}

\author{F. Peano$^{1,2}$}\email{fabio.peano@polito.it}\author{R. A. Fonseca$^{2}$}\author{J. L. Martins$^{2}$}\author{L. O. Silva$^{2}$}\email{luis.silva@ist.utl.pt}
\affiliation{$^1$Dipartimento di Energetica, Politecnico di Torino, corso Duca degli Abruzzi 24, 10129 Torino, Italy}\author{ }\affiliation{$^2$GoLP/Centro de F\'{\i}sica dos Plasmas, Instituto Superior T\'ecnico, 1049-001 Lisboa, Portugal}

%\author{F. Peano}
%\altaffiliation[Permanent Address: ]{I.N.F.M. and Dipartimento di Energetica, Politecnico di Torino, corso Duca degli Abruzzi 24, 10129 Torino, Italy}
%\email{Fabio.Peano@polito.it}
%\author{R. A. Fonseca}
%\author{L. O. Silva}
%\email{luis.silva@ist.utl.pt}
%\affiliation{GoLP/Centro de F\'{\i}sica dos Plasmas, Instituto Superior T\'ecnico, 1049-001 Lisboa, Portugal}

\date{\today}

\begin{abstract}
The ion phase-space dynamics in the Coulomb explosion of very large ($\sim 10^6 - 10^7$ atoms) deuterium clusters can be tailored using two consecutive laser pulses with different intensities and an appropriate time delay. For suitable sets of laser parameters (intensities and delay), large-scale shock shells form during the explosion, thus highly increasing the probability of fusion reactions within the single exploding clusters. In order to analyze the ion dynamics and evaluate the intracluster reaction rate, a one-dimensional theory is used, which approximately accounts for the electron expulsion from the clusters. It is found that, for very large clusters (initial radius $\sim$ 100 nm), and optimal laser parameters, the intracluster fusion yield becomes comparable to the intercluster fusion yield. The validity of the results is confirmed with three-dimensional particle-in-cell simulations.
\end{abstract}

%\pacs{Valid PACS appear here}% PACS, the Physics and Astronomy
                              % Classification Scheme.
%\keywords{Suggested keywords}% Use showkeys class option if keyword
                              % display desired

\maketitle

\section{Introduction}
\label{chap:intro}

Nowadays, CPA (chirped pulse amplification) laser technology \cite{Strickland} allows the table-top production of ultra-short ($\sim 10-100$ fs) laser pulses, with peak power up to the Petawatt level \cite{Perry}. When focusing such lasers to spot sizes of a few tens of micrometers, peak intensities up to $10^{21}$ W/cm$^2$ can be achieved, opening new research realms in the field of light-matter interaction. Among these, the interaction of ultra-intense lasers with clustered gases has become a central research topic \cite{Krainov_rep_1, Ditmire_Nature,Zweiback_PRL,Ditmire_POP_1,Grillon_PRL_1, electron_heating, Ditmire_PRL_3, Kim_PRL_1, Last_PRL_1}, mainly owing to the amazingly efficient coupling of ultra-intense laser radiation to clustered media; in experiments, nearly 100\% of the total laser energy has been observed to be deposited within a few millimeters propagation length \cite{Ditmire_PRL_3,Boyer}. This effective energy absorption results in various experimental evidences, such as bright x-ray emission \cite{xrays,Dobosz,Ditmire_PRL_4,Parra_2}, production of highly-ionized matter \cite{Ditmire_PRA_1,Jungreuthmayer_PRL_1,Koller}, generation of energetic electrons and ions \cite{Shao,Chen,Springate,Taguchi_PRL_1}, and copious production of fusion neutrons \cite{Ditmire_Nature,Zweiback_PRL,Ditmire_POP_1,Grillon_PRL_1}.

Depending on the gas, the size of the clusters, and the laser features, a huge variety of physical scenarios is possible, from the slow, hydrodynamic expansion of quasi-neutral nanoplasmas to the violent Coulomb explosion of highly-charged ion clouds.
In this paper, the attention is focused on the nonlinear ion dynamics in the Coulomb explosion of large ($10^6 - 10 ^7$ atoms) deuterium clusters irradiated with ultra-intense ($10^{18}-10^{21}$ W/cm$^2$) lasers. In the conditions considered here, all the atoms in a cluster are immediately stripped of their only electron, via field ionization, at the leading edge of the laser pulse, and the nanoplasma approximation \cite{Ditmire_PRA_1} can be adopted. The newly-formed free electrons can then be expelled, partially or totally, from the host cluster (outer ionization). If this process is extremely fast with respect to the explosion time scale, its dynamics may be neglected, but such an extreme scenario is met only with small clusters (less than 10000 atoms). For large clusters ($10^6 -10^8$ atoms), the outer ionization dynamics is crucial \cite{Peano,Madison_PRA_1,Parks} and it highly affects the explosion features.
In the present work, we explore the possibility of using two sequential laser pulses (``double pump'' henceforth) to control the outer ionization process and to drive large-scale shock shells \cite{Peano,Kaplan_PRL_2} during the Coulomb explosion of large deuterium clusters. As a consequence of the shock shell formation, relative velocities appear between ions belonging to a single exploding cluster and intracluster DD fusion reactions may take place in the early phase of the explosion, long before the neighboring clusters start interacting with one another. It is found that, for optimal double-pump parameters, the intracluster fusion yield can become comparable with the intercluster fusion yield. Hence, for appropriate experimental conditions, a time-resolved burst of fusion neutrons should be detected before the usual bulk of fusion neutrons, thus providing a clear experimental evidence for the formation of shock shells on a nanometer scale \cite{Kaplan_PRL_2}.

In the present work, the laser-induced Coulomb explosion is modeled using an approximate 1D theory, in which a prescribed electron dynamics, related to the pulse envelope function, is assumed \cite{Peano}. The validity of the results is then confirmed with highly-realistic, three-dimensional (3D) particle-in-cell (PIC) \cite{langdon}. The simulation framework adopted here is OSIRIS 2.0 \cite{osiris}, a state-of-the-art, massively-parallel, electromagnetic, fully-relativistic, 3D PIC code.

This paper is organized as follows. In Section \ref{sec:shock_shells}, the formation of large-scale shock shells in Coulomb explosions is introduced. In Section \ref{sec:1D_model}, a 1D model for the Coulomb explosion is outlined, with an emphasis on the inclusion of the effects caused by the electron dynamics. In Section \ref{sec:intracluster}, the 1D model is employed to study the intracluster fusion reactions. The fusion yield from intracluster and intercluster reactions are compared in Section \ref{sec:ic_vs_IC}, and three-dimensional PIC simulations of the double-pump technique are presented in Section \ref{sec:3D_sim}. Finally, the conclusions are stated.

\section{Large-scale shock shells in Coulomb explosions}
\label{sec:shock_shells}

The possible formation of shocks \cite{shocks} during the Coulomb explosion of an ideal pure-ion sphere has been recently demonstrated theoretically \cite{Kaplan_PRL_2}. In particular, it has been shown that every radial nonuniformity in the initial ion density profile leads to the formation of a multi-branch phase-space structure, which is named a shock shell, accompanied by the appearance of one or more peaks in the ion density profile. Such structures form when the initial ion density is nonuniform and a maximum in the radial velocity profile soon forms at some radius within the distribution, leading to a characteristic overtaking process between the ions. 

One interesting consequence of the shock formation is the possible occurrence of energetic ion-ion collisions within a single expanding cluster. For instance, the appearance of pronounced shock shells in the explosion of very large ($10^{6} - 10^{7}$ atoms) deuterium clusters could cause an enhancement in the probability of intracluster fusion reactions. According to the theory, in order to obtain a large-scale shock shell, the nonuniformity in the initial density profile must be large-scale as well. In fact, if the initial density profile differs just slightly from a uniform, step-like profile, the corresponding shock shell is doomed to be small-scale and cannot involve significant relative velocities and collision energies.
Results from 3D numerical simulations of the interaction of an ultra-intense laser with large deuterium clusters in realistic conditions showed that shocks may form naturally even when starting from uniform, step-like density profiles \cite{Peano} because the electron dynamics in the laser field smooths out the ion density profile in the early stages of the interaction.
However, the simulations also showed these spontaneously-occurring shocks are small-scale, so that no appreciable relative velocities appear within the cluster, which means that no shock-driven, intracluster fusion events occur.

\begin{figure}[ht]
\epsfig{file=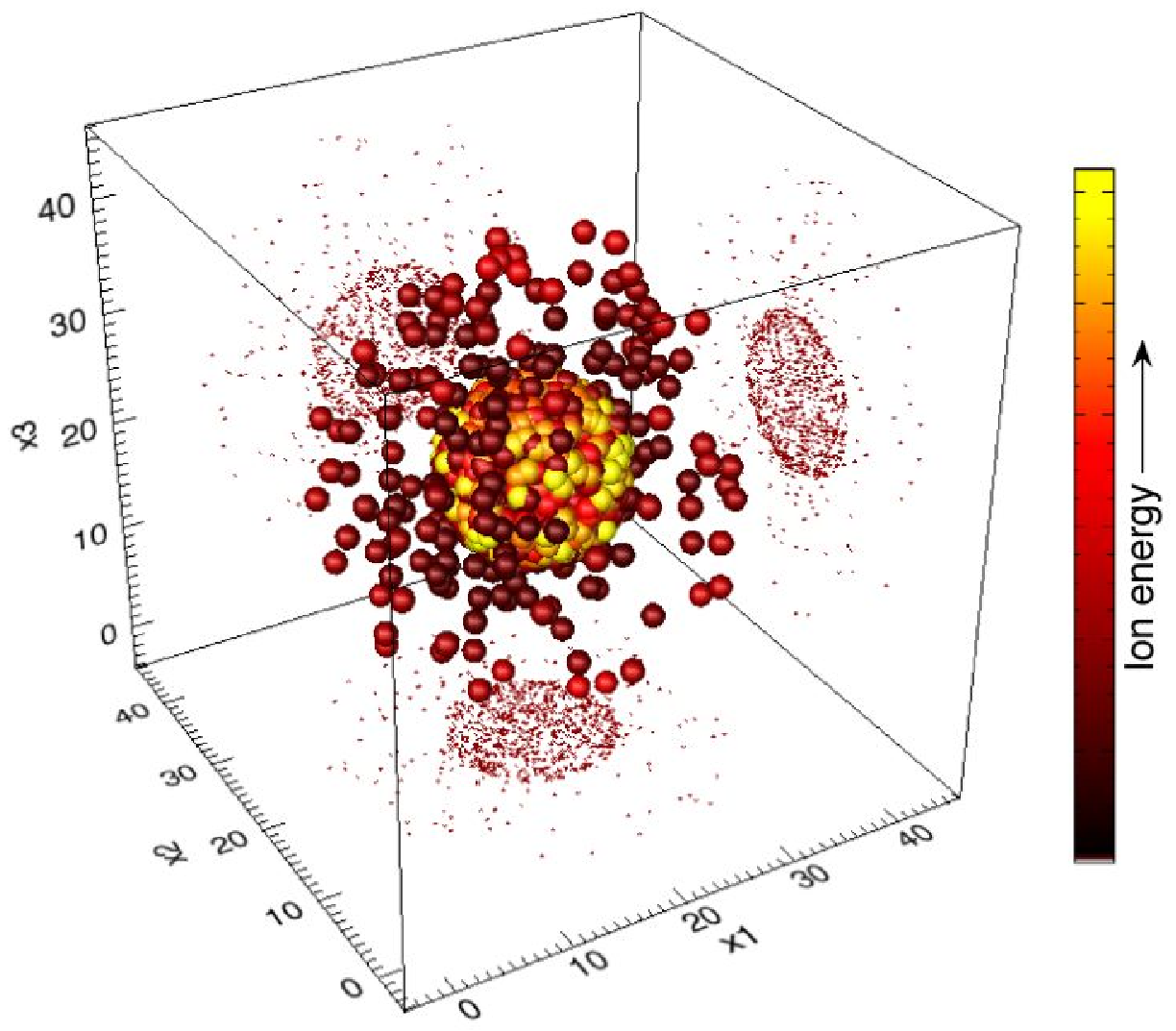, width=7cm}
\caption{(Color online) Ion distribution in configuration space at time $t=350$ fs for the double-pump case described in Section \ref{sec:3D_sim}. The colored spheres indicate position and energy of a random sample of $\sim 1 \times 10^3$ (out of $\sim 6.4 \times 10^6$) particles. Color is proportional to energy, the lightest spheres corresponding to the maximum ion energy, $\mathcal{E}\ped{max}=120$ keV.}
\label{fig:explosion}
\end{figure}

At first sight, inducing such reactions via generation of large-scale shock shells in exploding deuterium clusters would then appear unfeasible, as it would require the ability to tailor the initial radial density profile of roughly spherical objects on the nanometer scale. Surprisingly, on the contrary, such an apparently-difficult goal can be achieved quite simply by using a suitable sequence of laser pulses with different intensities \cite{Peano}. In fact, if the first pulse is relatively weak, only a small number of electrons are stripped off the host cluster, and a slow expansion takes place, driven by both Coulomb repulsion and hydrodynamic pressure of the hot electrons \cite{Ditmire_PRA_1}. As the expansion goes on, the ion density in the outer region of the cluster decreases, while the cluster core remains dense and approximately neutral: in this way, a smoothly-decreasing plasma density profile is naturally formed. Afterwards, if the electrons are suddenly swept away from the cluster core by
a second, extremely-intense pulse, the inner ions immediately feel a much higher repulsive force than the outer ions do. The cluster core explodes abruptly and the fastest inner ions overrun the slowly-expanding outer ions, leading to the formation of a large-scale shock shell. Such a situation is well depicted in Fig. \ref{fig:explosion}, where colored spheres indicate position and energy of a random sample of ions, after the passage of the second pulse (the figure has been built from the results of the 3D PIC simulation discussed in Section \ref{sec:3D_sim}).

The double-pump technique not only provides an effective method to induce large-scale shock shells in the Coulomb explosion of very large clusters, but it also provides the ability to control the phase-space structure of the exploding clusters, by varying the delay between the two pulses, $\Delta t$, and the peak intensity of the first pulse, $I_{1}$. The dependence of the explosion features on these parameters is analyzed in detail in the following sections, where we investigate the possibility of inducing intracluster, shock driven fusion reactions in double-pump experiments. 

Alternatively, large-scale shock shells can also be driven using clusters constituted by more than one species, as long as the ion charge-to-mass density ratio decreases radially. Such possibility will be explored in a future publication \cite{Martins}.
 
\section{1D model for Coulomb explosion}
\label{sec:1D_model}

A proper tuning of the double-pump parameters requires a deep knowledge of their influence on the explosion dynamics, the shock shell formation and evolution, and, consequently, the intracluster fusion yield. Since a direct parameter scan via either 2D or 3D PIC simulations would have been impracticable, because it is too computationally demanding, we have developed a simple 1D theoretical model that allowed us to investigate the effects of the key double-pump parameters, namely the time delay between the pulses, $\Delta t$, and the peak intensity of the first pulse, $I_{1}$.

Our model, which is an extended version of the standard model for the Coulomb explosion of a spherical pure-ion cluster (cf. \cite{Kaplan_PRL_2}), is suitable for both spherical and cylindrical symmetry and takes into account the effect of the electron population in the expanding cluster. The electron population follows a prescribed dynamics, determined uniquely by the laser pulse features. In this framework, the acceleration of an ion at a given time $\tau$ and radial position $r$ can be written, in dimensionless units, as  
\begin{equation}
\frac{d^2r}{d\tau^2} = \frac{ Q( r ) - Q\ped{el}( r , \tau ) } {r^{\gamma}} \text{,}
\label{eq:acc1}
\end{equation}
where ${\gamma}$ accounts for the geometry (${\gamma}=2$ for spherical geometry, ${\gamma}=1$ for cylindrical geometry), and where mass is normalized to $m$ (ion mass), length to $R_{0}$ (initial radius of the cluster), charge to the elementary charge $e$, and time to $t_0=\sqrt{ \left(m R_0^{\gamma+1}\right)/\left(e^2 N_{0}\right)}$; $t_0$ also represents the time scale for the explosion of a pure-ion spherical/cylindrical distribution, being $N_{0}$ the total number of ions (per unit length in the cylindrical case). The quantity $Q(r)$ is the ion charge within a sphere/cylinder of radius $r$, while $Q\ped{el}( r , \tau )$ describes the absolute value of the electron charge still present within the same sphere/cylinder at time $\tau$.
Using the Cluster Barrier Suppression Ionization (CBSI) model described in \cite{Last_JCP_1}, $Q\ped{el}( r , \tau )$ can be determined from the instantaneous value of the laser-field envelope function, $E_l(\tau)$. According to the CBSI theory, an electron is expelled from the cluster boundary to infinity whenever
\begin{equation}
\frac{E_l(\tau)}{\sqrt{2}} \geqslant  \frac{ 1 - Q\ped{el}(R, \tau ) } {R^{\gamma}} \text{,}
\label{eq:oution1}
\end{equation}
where $R$ is the cluster radius at time $\tau$ (the $\sqrt{2}$ factor accounts for the periodicity of the laser electric field). To determine $Q\ped{el}( r , \tau )$ we need a further assumption on the radial profile of the electron distribution. The simplest choice is to assume that, at each time $\tau$, the remaining electrons fully neutralize the cluster core. The permanence of a core of cold electrons within the cluster is clearly visible in PIC simulations \cite{Peano,Kishimoto_POP_1} and it has been recently explained theoretically in \cite{Breizman}. Thus, $Q\ped{el}( r , \tau )$ takes the form
\begin{equation}
Q\ped{el}(r,\tau) = \left\{
\begin{array}{ccc}
Q(r) &  & r \leqslant R\ped{el}(\tau) \\Q(R\ped{el}(\tau), \tau) &  & r > R\ped{el}(\tau)
\end{array}
\right. \text{,}
\label{eq:Qel}
\end{equation}
being
\begin{equation}
R\ped{el}(\tau) = \left[1 -  R^{\gamma}\frac{E_l(\tau)}{\sqrt{2}} \right]^{\frac{1}{\gamma+1}}
\label{eq:Rel}
\end{equation}
the radius of the electron sphere/cylinder at time $\tau$.
Under these assumptions, Eq. (\ref{eq:acc1}) can be written as
\begin{equation}
\frac{d^2r}{d\tau^2} = \left\{
\begin{array}{ccc}
0 &  & r \leqslant R\ped{el}(\tau) \\
\dfrac{ Q( r ) - \left[ 1 - R^{\gamma}E_l(\tau)/\sqrt{2} \right]} {r^{\gamma}} &  & r > R\ped{el}(\tau)
\end{array}
\right. \text{,}
\label{eq:acc2}
\end{equation}
providing a simple model for studying the Coulomb explosion of large clusters driven by a general sequence of laser pulses, having different intensities and envelopes. This model can be used to analyze scenarios involving the formation of large-scale shock shells, allowing the control of the shock features through tuning of the pulse parameters.

As long as the initial ion density profile $\rho_0(r_0)$ is known, Eq. (\ref{eq:acc2}) can be readily integrated numerically by following the trajectory $r(r_0,\tau)$ of a finite set of ions with different initial position $r_0$. Details of the numerical model will be described elsewhere.
 
\section{Intracluster fusion reactions}
\label{sec:intracluster}

\begin{figure}[ht]
\epsfig{file=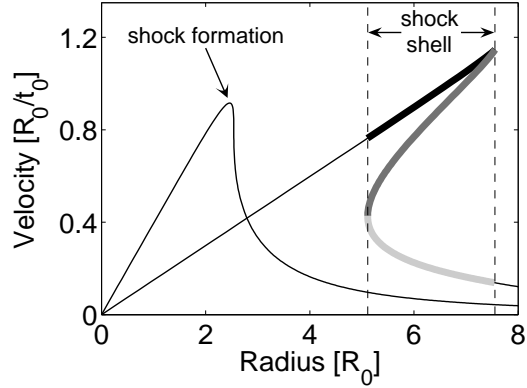, width=7cm}
\caption{Typical three-branched shock shell in phase space (thick, gray lines mark the different branches) \cite{Peano, Kaplan_PRL_2} for the Coulomb explosion of a pure-ion sphere with nonuniform radial density profile. Units are normalized as in Section \ref{sec:1D_model}.}
\label{fig:shock_shell}
\end{figure}

Large-scale shock shells are characterized by a well-defined multi-branch structure in the $v-r$ phase space, most frequently a three-branch structure \cite{Kaplan_PRL_2, Peano} as the one in Fig. \ref{fig:shock_shell}, which refers to the Coulomb explosion of a pure-ion sphere with a nonuniform radial density profile \cite{Kaplan_PRL_2}. As the explosion goes on, the upper branch overlaps the lower branches: the shock shell widens radially, narrowing its velocity spread, and, meanwhile, the ion density on each branch decreases. Therefore, one can reasonably expect the probability of nuclear reactions between ions belonging to different branches to be higher in the early stages after the shock shell formation and to decrease rapidly at advanced times.

At each radius $r$ and time $\tau$, the number of reactions per unit time and unit volume, $\mathcal{R}$, is given (in dimensionless units) by
\begin{equation}
\mathcal{R} = \frac{1}{2}\int_{v}\int_{v^{\prime}}f\left( r, v\right)f\left( r, v^{\prime}\right) \hat{\sigma}\left( | v -  v^{\prime} |  \right) | v -  v^{\prime} | \diff{v} \diff{v^{\prime}} \text{,}
\label{eq:rrate1}
\end{equation}
where $f\left( r, v\right)$ is the 1D distribution function for the ions and $\hat{\sigma}$ the normalized DD fusion cross section ($\hat{\sigma} = \sigma/R_0^2$, with $\sigma$ in cm$^2$). Outside the shock shell, this integral vanishes because there are no relative velocities; within the shock shell, it simply reduces to the sum over the three branches of the phase space profile (identified in color in Fig. \ref{fig:shock_shell}):
\begin{equation}
\mathcal{R} = \frac{1}{2} \sum\ped{i,j=1}^{3}
                         \rho\ped{i}(r)
                         \rho\ped{j}(r)
                         \hat{\sigma} \left( | v\ped{i} - v\ped{j}|  \right)
                         |v\ped{i} - v\ped{j}|
\label{eq:rrate2}
\end{equation}
where $\rho\ped{i}(r) = 1/\left(2\gamma \pi {r}^{\gamma}\right)\partial Q\ped{i}/\partial r$ is the ion density on the ith branch.
The intracluster reaction rate, $\boldsymbol{R}$, is then
\begin{equation}
\boldsymbol{R} =  N^{2}_{0} \int_{r\ped{sh}}^{R\ped{sh}}
              \mathcal{R} \
              2\gamma \pi {r}^{\gamma}
              \diff{r} \text{,}
\label{eq:rrate3}
\end{equation}
where $r\ped{sh}$ and $R\ped{sh}$ represent the shock shell boundaries. The number of reactions per cluster, $\mathcal{N}$, is given by
\begin{equation}
\mathcal{N} = \int_{\tau\ped{sh}}^{\infty}\boldsymbol{R}
              \diff{\tau}
\label{eq:nreact1}
\end{equation}
where $\tau\ped{sh}$ is the shock formation time. Once the phase space history of each ion of the initial distribution is known, from the solution of Eq. (\ref{eq:acc2}), $\mathcal{N}$ can be evaluated numerically through Eqs. (\ref{eq:rrate2})-(\ref{eq:nreact1}). 

The 1D theory here described provides a useful framework to perform parametric studies and investigate the influence of $I_{1}$ and $\Delta t$ on the total number of reactions per cluster, $\mathcal{N}$, seeking the combination of parameters that maximizes it: a maximum in $\mathcal{N}$ is expected to show up when the first pulse is intense enough to drive the first expansion but not so intense to expel too many electrons from the cluster. The delay of the second pulse needs to be long enough to allow the formation of a decreasing density profile but not so long to let the outer ions expand to large $r$, far from the cluster core.

When analyzing the results of the parametric studies, the approximations introduced in the 1D theory must be kept in mind. First, as hydrodynamic effects are neglected, the effect of the first pulse is underestimated when dealing with low values of $I_{1}$, meaning that, in reality, the first expansion is faster than what is predicted by Eq. (\ref{eq:acc2}) and the optimal value of $I_{1}$ is actually lower than expected. Furthermore, the 1D theory cannot include laser polarization effects which may affect the dynamics by causing an asymmetry in the explosion. Finally, the 1D theory with CBSI model is expected to break down with enormous clusters ($R_0 > 500$ nm - 1 $\mu$m), having initial radius similar to the laser central wavelength, meaning that the predicted optimal double-pump parameters could be unreliable in such cases.

\begin{figure}[ht]
\epsfig{file=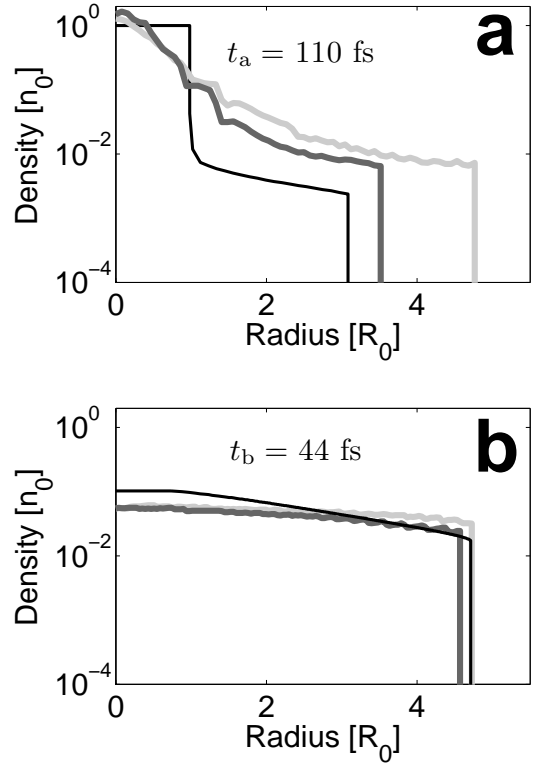, width=7cm}
\caption{Ion number density when the cluster radius is $\sim 5R_0$: (a) low-intensity case, and (b) high-intensity case. Thick gray lines represent lineouts in the $\hat{x}$ (dark) and $\hat{y}$ (light) directions. Thin black lines refer to the solution obtained from the 1D theoretical model.}
\label{fig:1D-vs-PIC}
\end{figure}

We have tested the validity of the reduced model by comparison with a series of results from single-pump 2D PIC simulations in various configuration of cluster size and laser peak intensity. Figures \ref{fig:1D-vs-PIC}a and \ref{fig:1D-vs-PIC}b show comparisons between lineouts (along the laser propagation ($\hat{x}$) and polarization ($\hat{y}$) directions) of the ion density distribution, taken from the simulation results, and the density profile predicted by the 1D theory. Both plots refer to the interaction of a circular, rodlike cluster (initial radius $R_0 = 32$ nm, particle density $n_{0}=4.56\times10^{22}$ $\text{cm}^{-3}$) with a laser pulse  having central wavelength $\lambda_0=820$ nm, and approximately Gaussian envelope with rise time $t\ped{rise}=35$ fs. The peak intensities are $I\ped{a} = 4 \times 10^{16}$ $\text{W}/\text{cm}^2$ (Fig. \ref{fig:1D-vs-PIC}a) and $I\ped{b} = 1.6 \times 10^{19}$ $\text{W}/\text{cm}^2$ (Fig. \ref{fig:1D-vs-PIC}b). As expected, the results from the 1D model are quantitatively more accurate in the high-intensity case, where Coulomb forces are dominant, than in the low-intensity case, where hydrodynamic effects and polarization effects are relevant, and the cluster expands more rapidly than what is predicted by the 1D theory. 
For these reasons, when exploring the possibility of intracluster fusion reactions, we resorted to the 1D model to seek a good combination of double-pump parameters, which we then adjusted and employed to perform accurate 3D PIC simulations presented in Section \ref{sec:3D_sim}.  
 
The model also provides information on how the initial cluster size $R_{0}$ affects $\mathcal{N}$, allowing one to perform PIC simulations using clusters with initial radius $R_{0} \sim 10$ nm and then to extrapolate the results to the case of extremely large clusters having initial radius $R_{0} \sim 100$ nm, without performing new, and extremely large, simulations.

\begin{figure}[ht]
\epsfig{file=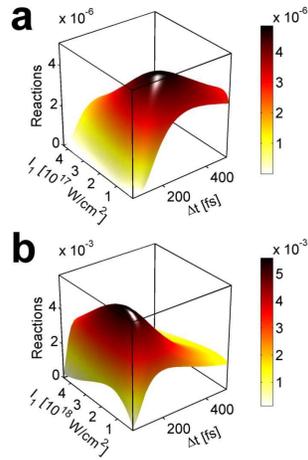, width=7cm}
\caption{(Color online) Total number of reactions per cluster, $\mathcal{N}$, as a function of the peak intensity of the first laser pulse, $I_{1}$, and the time delay between the pulses, $\Delta t$, for (a) $R_0= 32 $ nm and (b) $R_0= 100 $ nm.}
\label{fig:DP_opt}
\end{figure}

\begin{figure}[ht]
\epsfig{file=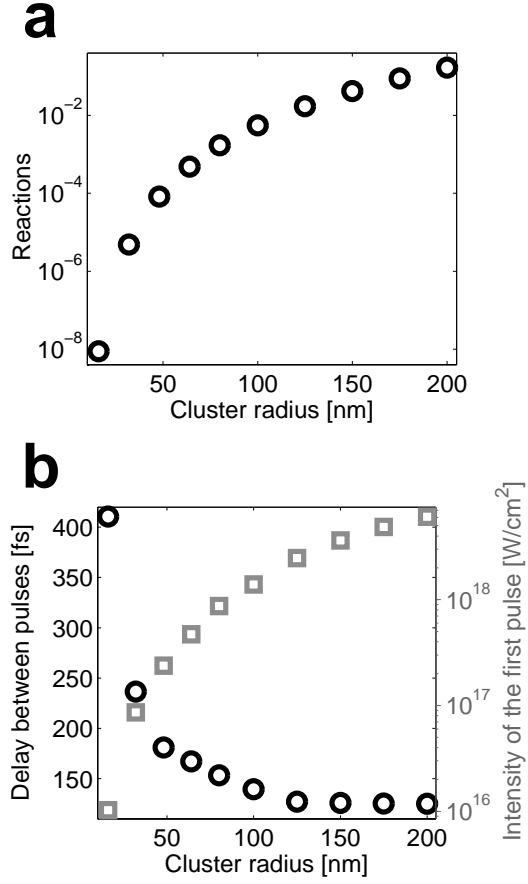, width=7cm}
\caption{(a) Maximum value of the total number of reactions per cluster, $\mathcal{N}\ped{max}$, and (b) optimal values for the double pump parameters $I_{1}$ (gray squares) and $\Delta t$ (black circles), for different values of the initial cluster radius, $R_0$.}
\label{fig:parametric}
\end{figure}

Here, we present results from parametric studies, with respect to $I_{1}$ and $\Delta t$, for several cluster sizes, with initial radii, $R_{0}$, in the range $16-200$ nm. For each $R_{0}$, we consider a spherical cluster of atomic deuterium (with uniform step-like density profile, $n_{0}=4.56\times10^{22}$ $\text{cm}^{-3}$) hit by a pulse sequence in which a weak laser pulse (peak intensity $I_{1}$ (variable), central wavelength $\lambda\ped{0,1} = 820$ nm, and approximately Gaussian envelope with rise time $t\ped{rise,1} \sim 35$ fs) is followed by an ultra-intense pulse (peak intensity $I_{2} \gg I_{1}$, central wavelength $\lambda\ped{0,2} = 820$ nm, and approximately-Gaussian envelope with rise time $t\ped{rise,2} \sim 20$ fs) with time delay $\Delta t$ variable in the range $70 - 500$ fs. In all cases, the peak intensity of the second pulse is assumed to be high enough to expel all the electrons from the cluster core and drive a sudden Coulomb explosion. Figure \ref{fig:DP_opt} shows the total number of reactions per cluster, $\mathcal{N}$, as a function of $I_{1}$ and $\Delta t$, for two representative cases: $R_{0} = 32$ nm (Fig. \ref{fig:DP_opt}a) and $R_{0} = 100$ nm (Fig. \ref{fig:DP_opt}b).
With $R_{0} = 32$ nm, $\mathcal{N}$ assumes its maximum value, $\mathcal{N}\ped{max} = 4.85 \times 10^{-6}$ reactions, for $I_{1} = 8.6 \times 10^{16}$ $\text{W}/\text{cm}^2$, $\Delta t = 236$ fs. With $R_{0} = 100$ nm, $\mathcal{N}\ped{max} = 5.58 \times 10^{-3}$ reactions for $I_{1} = 1.4 \times 10^{18}$ $\text{W}/\text{cm}^2$, $\Delta t = 139$ fs. Hence, approximately trebling the cluster size (from $R_{0} = 32$ nm to $R_{0} = 100$ nm) results in gaining three order of magnitudes in $\mathcal{N}$, with a first pulse sixteen times as intense and a much shorter delay. These drastic changes are partly due to the variation of the DD fusion cross section, $\sigma$, with the collision energy. In fact, as follows from Eqs. (\ref{eq:rrate2})-(\ref{eq:nreact1}), if $\sigma$ were constant, $\mathcal{N}\ped{max}$ would be proportional to $R^{4}_{0}$, the optimal intensity for the first pulse would scale as $R^{2}_{0}$, while the optimal delay would stay the same. In that case, the two plots in Fig. \ref{fig:DP_opt} would have the same shape and $\mathcal{N}\ped{max}$ would increase by a factor less than 100. The dependence of the optimal combination of $I_{1}$ and $\Delta t$ on the initial cluster size is depicted in Fig. \ref{fig:parametric}, along with the corresponding variation of $\mathcal{N}\ped{max}$. As $R_{0}$ increases, the optimal intensity of the first pulse increases as well, while the optimal delay decreases and appears to saturate towards the value $\Delta t = 125$ fs.

\begin{figure}[ht]
\epsfig{file=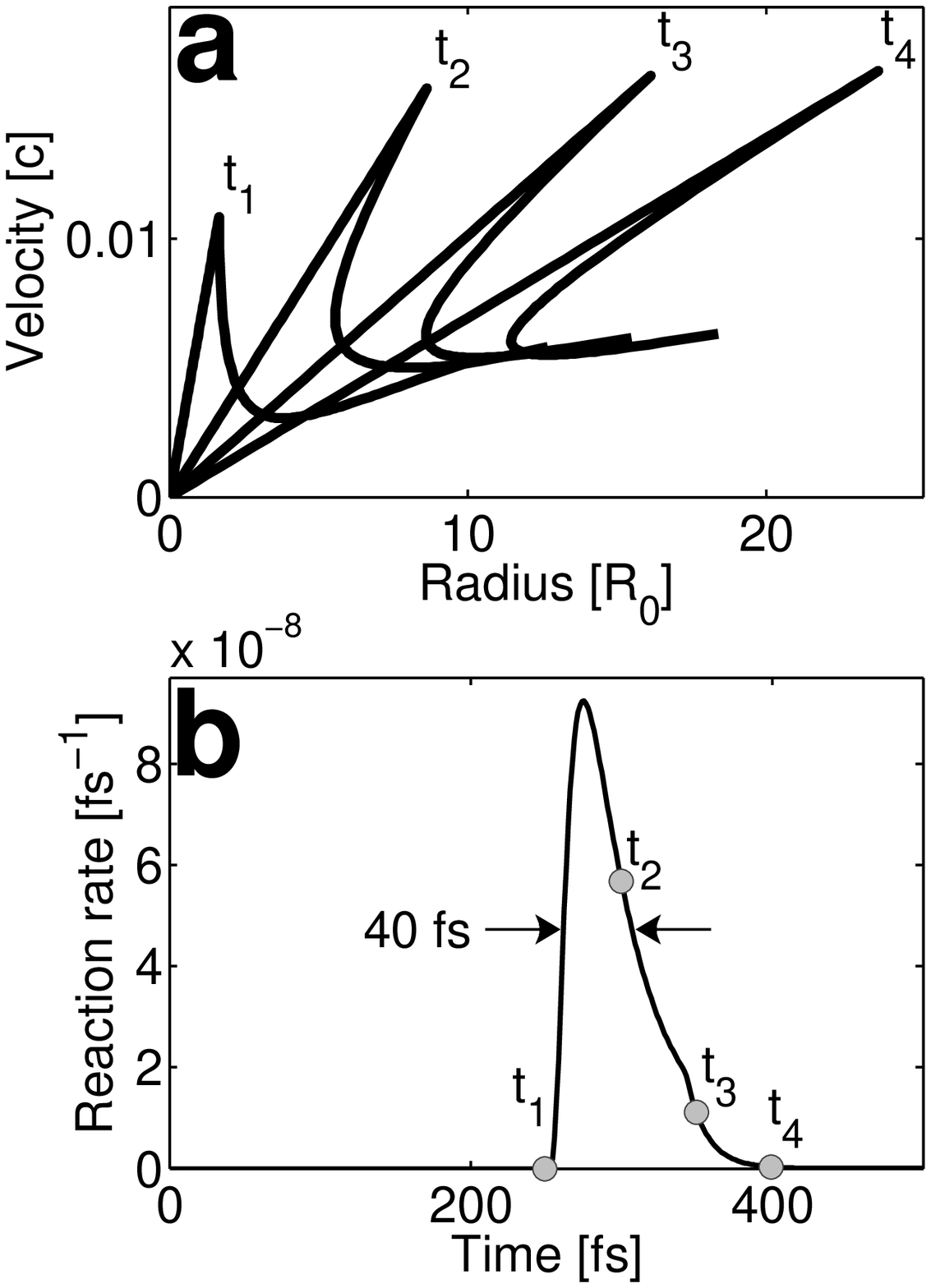, width=7cm}
\caption{(a) Phase space profile at times $t_1 = 255$ fs, $t_2 = 305$ fs, $t_3 = 355$ fs, and $t_4 = 405$ fs; (b) time history of the reaction rate, $\boldsymbol{R}$, for $R_0= 32 $ nm, and with the optimal combination of double pump parameters: $I_{1} = 8.6 \times 10^{16}$ $\text{W}/\text{cm}^2$, $\Delta t = 236$ fs.}
\label{fig:dynamics_1}
\end{figure}

\begin{figure}[ht]
\epsfig{file=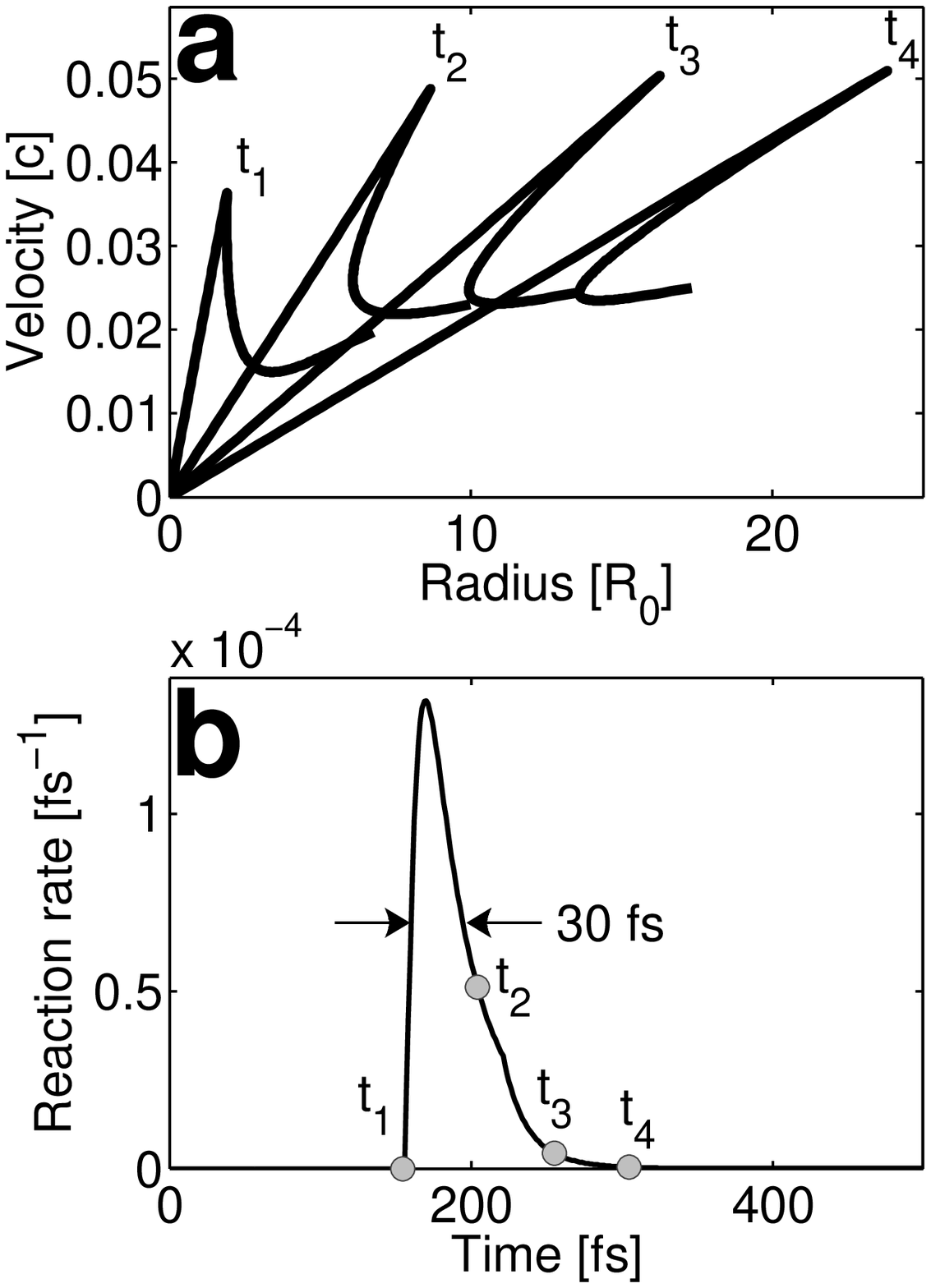, width=7cm}
\caption{(a) Phase space profile at times $t_1 = 155$ fs, $t_2 = 205$ fs, $t_3 = 255$ fs, and $t_4 = 305$ fs; (b) time history of the reaction rate, $\boldsymbol{R}$, for $R_0= 100 $ nm, and with the optimal combination of double pump parameters: $I_{1} = 1.4 \times 10^{18}.$ $\text{W}/\text{cm}^2$, $\Delta t = 139$ fs.}
\label{fig:dynamics_2}
\end{figure}

For both cases showed in Fig. \ref{fig:DP_opt}, we report the evolution of the $v-r$ phase space profile (starting right after the second pulse reaches the expanding cluster causing the formation of the shock shell), along with the time history of the reaction rate $\boldsymbol{R}$, in Figs. \ref{fig:dynamics_1} and \ref{fig:dynamics_2}, respectively. As one can see, $\boldsymbol{R}$ exhibits a sharp peak immediately after the shock formation, which occurs at time $t_{1}$, causing the probability of intracluster fusion reactions to retain appreciable values only for a time interval of a few tens of fs. Intracluster fusion reactions represent a much faster, and profoundly different, phenomenon than intercluster reactions (whose typical time scale is $\sim 10-100$ ps \cite{Last_PRA_1}): they occur abruptly in the single clusters when the particular phase-space dynamics origins a tiny, high-density, and short-lived reaction volume (the shock shell). On the contrary, intercluster reactions occur later in the big, long-lived reaction volume which is the whole plasma filament created by the laser pulses. An estimate of the respective contributions of the two, distinct processes to the total fusion yield is presented below. 

\section{Intracluster and intercluster fusion yields}
\label{sec:ic_vs_IC}

In the last years, various experiments revealed the occurrence of nuclear fusion reactions in clustered media irradiated by ultra-intense fs lasers \cite{Ditmire_Nature,Zweiback_PRL,Ditmire_POP_1,Grillon_PRL_1}. Most theoretical models developed to explain the experimental results showed that fusion reactions arise primarily from collisions between fast ions belonging to different clusters  \cite{Ditmire_Nature,Zweiback_PRL,Ditmire_POP_1,Grillon_PRL_1,Parks,Last_PRA_1,Kishimoto_POP_1}, though the contribution from collisions of fast ions with colder ions and atoms outside the plasma filament created by the laser has also been investigated \cite{Madison_POP_1}. Here, we analyze the role of intracluster, shock-driven fusion reactions in double-pump scenarios, for different cluster sizes, and we compare the intracluster fusion yield with the intercluster fusion yield.

For the calculation of the intercluster fusion yield, $Y\ped{IC}$, we refer to the simple model presented in \cite{Last_PRA_1}, where $Y\ped{IC}$ is evaluated as
\begin{equation}
Y\ped{IC} = \frac{1}{2}\bar{n}^{2}\left\langle \sigma v\right\rangle V\ped{\textbf{r}} T\ped{d}
\label{eq:Y_intercl}
\end{equation}
being $\bar{n}$ the average atomic density inside the reaction volume $V\ped{\textbf{r}}$, $T\ped{d}$ the plasma disassembly time (typical time for the expansion of the plasma in the reaction volume), and
\begin{eqnarray}
\left\langle \sigma v\right\rangle &=& \frac{R_0}{2t_0}\int^{\infty}_{0} \!\!\!\! \int^{\infty}_{0} \!\!\!\! \int^{\pi}_{0}
P\left(\mathcal{E}_{1}\right)P\left(\mathcal{E}_{2}\right) \sigma(\mathcal{E}\ped{coll}) \times \nonumber \\
& & \times \left(2\mathcal{E}\ped{coll}/m \right)^{1/2} \sin\left(\alpha \right)  \diff{\mathcal{E}_{1}}\diff{\mathcal{E}_{2}}\diff{\alpha}
\label{eq:sigmav_1}
\end{eqnarray}
where $\mathcal{E}\ped{coll} = \mathcal{E}_{1}+\mathcal{E}_{2}-2\left(\mathcal{E}_{1}\mathcal{E}_{2}\right)^{1/2}\cos(\alpha) $ is the binary collision energy for particles with kinetic energy $\mathcal{E}_{1}=v_1^2/2$, $\mathcal{E}_{2}=v_2^2/2$, and collision angle $\alpha$. The quantity $P\left(\mathcal{E}\right)$ is the (dimensionless) energy distribution of a single exploding cluster, calculated as
\begin{eqnarray}
P\left(\mathcal{E}\right) &=&  \int^{\infty}_{0}\!\!\!\!\int^{\infty}_{0}f\left(r^{\prime},v^{\prime}\right)\delta\left({v^{\prime}}^2/2-\mathcal{E}\right)4\pi {r^{\prime}}^2 \diff r^{\prime}\diff v^{\prime} = \nonumber \\ 
&=& \sum^{N\ped{z}}_{j=1}\frac{1}{v\ped{j}}\left.\frac{dQ}{dv}\right|_{v = v\ped{j}}
\label{eq:spectrum}
\end{eqnarray}
being $\{v\ped{j}\}$ the $N\ped{z}$ zeros of $v^2/2-\mathcal{E}$. The reaction volume is assumed to be a cylinder with radius $R\ped{\textbf{r}}=100$ $\mu\text{m}$ and height $H\ped{\textbf{r}}=2$ $m\text{m}$, and the plasma disassembly time is estimated as $T\ped{d} \sim \pi R\ped{\textbf{r}}/\left(2v\ped{max}\right)$ \cite{Last_PRA_1}.

\begin{figure}[ht]
\epsfig{file=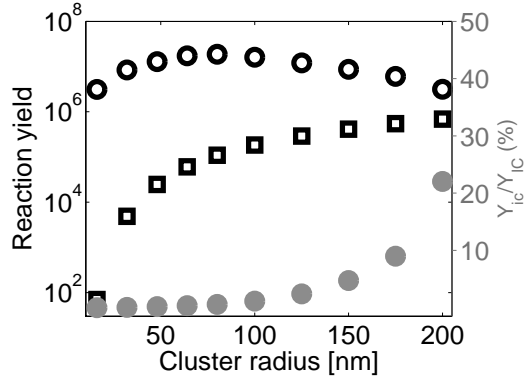, width=7cm}\caption{Intercluster fusion yield, $Y\ped{IC}$ (black circles), intracluster fusion yield, $Y\ped{ic}$ (black squares), and percentage value of $Y\ped{ic}/Y\ped{IC}$ (gray bullets) for different values of the initial cluster radius, $R_0$.}
\label{fig:yields}
\end{figure}

\begin{figure}[ht]
\epsfig{file=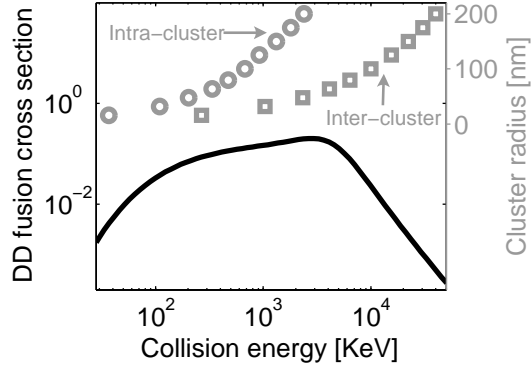, width=7cm}\caption{Cross section for the deuterium-deuterium fusion reaction, $\sigma$, as a function of the collision energy (full, black line); gray squares and gray circles indicate, respectively, the maximum intercluster and intracluster collision energy for different values of the initial cluster radius, $R_0$ (reported on the right, gray axis).}
\label{fig:sigma}
\end{figure}

In the previous section, we analyzed the explosion dynamics of a single cluster, neglecting its interaction with the neighboring clusters, and calculated the probability of intracluster reactions. Now, and in order to compare intracluster and intercluster fusion yields, we must consider the whole spatial distribution of clusters. For the sake of simplicity, we consider the ideal situation of a reaction volume containing a fixed number of deuterium atoms grouped in  $N\ped{cl}$ spherical clusters of equal size $R_0$, with a mean separation distance $d$, distributed on a regular square lattice. Given the average atomic density $\bar{n}$, the total number of clusters is $N\ped{cl} = \pi R\ped{\textbf{r}}^{2} H\ped{\textbf{r}} \bar{n}/N_0$ and the intercluster distance is $d = R_0\left[4\pi n_0/\left(3\bar{n}\right)\right]^{1/3}$. The intracluster fusion yield, $Y\ped{ic}$, can be evaluated as
\begin{equation}
Y\ped{ic}=N\ped{cl}\mathcal{N}
\label{eq:Y_intracl}
\end{equation}
provided that the ratio $\delta = d/R_0$ (which depends only on the packing fraction $\bar{n}/n_0$ \cite{Kishimoto_POP_1}) be sufficiently high ($\delta > 20-25$) and, consequently, most intracluster reactions occur before the cluster cores start interacting with one another. In the opposite case (small values of $\delta$: $\delta < 10-15$), almost no intracluster reactions can occur before the cluster cores start interacting with one another. This is explained by noticing that, typically, the intracluster reaction rate, $\boldsymbol{R}$, peaks when the radius of the outer boundary of the shock shell, $R\ped{sh}$, is $R\ped{sh} \sim 4-5 R_0$, and decreases below 1/10 the peak value when $R\ped{sh} \sim 10-15 R_0$ (see Figs. \ref{fig:dynamics_1}, \ref{fig:dynamics_2}).
If $\bar{n}=10^{19}$ $\text{cm}^{-3}$, a typical experimental value, and with the cluster density $n_{0}=4.56\times10^{22}$ $\text{cm}^{-3}$ here considered, one obtains $d/R_0 \simeq 27$. In such a low packing fraction case, $Y\ped{ic}$ can be evaluated through Eq. (\ref{eq:Y_intracl}) and a comparison with $Y\ped{IC}$ is readily carried out. Figure \ref{fig:yields} reports the value of $Y\ped{IC}$ and $Y\ped{ic}$ for different cluster sizes: for each value of $R_0$ we sought the optimal combination of double-pump parameters (see Fig. \ref{fig:parametric}) and then calculated the corresponding fusion yields.
The contribution of the intracluster reactions grows rapidly with $R_0$, meaning that the growth of the shock shells and the increase of the fusion cross section prevail against the decrease of the number of clusters in the reaction volume. On the contrary, the intercluster fusion yield, though keeping a high value, decreases at very high $R_0$ because collision energies beyond the one that maximizes $\sigma$ start to appear. This is well illustrated in Fig. \ref{fig:sigma}, where the maximum intracluster and intercluster collision energies are compared for different cluster sizes, along with $\sigma\left(\mathcal{E}\ped{coll}\right)$. For initial cluster radii $\gtrsim 70-80$ nm, the maximum collision energy for intercluster reactions lies to the right of the peak in $\sigma\left(\mathcal{E}\ped{coll}\right)$, while the maximum collision energy for intracluster reactions lies to the left of it even for initial cluster radii as high as $200$ nm: this explains the results of Fig. \ref{fig:yields}.

From our discussion, we conclude that, in principle, a double pump experiment with optimized pulse parameters, very-large clusters, and low packing fraction, should provide a clear signature for the occurrence of intracluster, shock-driven fusion reactions, in the form of a time-resolved burst of fusion neutrons which anticipates the bulk of fusion neutrons produced via both intercluster reactions within the plasma filament and ion-ion/ion-atom collisions outside the plasma filament.

\section{Three dimensional PIC simulations}
\label{sec:3D_sim}

In order to check the validity of the analysis presented above, and to get deeper physical insights, we have performed 3D PIC simulations of the laser-cluster interaction in a double-pump case, treating self-consistently the dynamics of electrons and ions in the laser field, the outer ionization dynamics, and the full dynamics of both the slow expansion induced by the first pulse and the sudden explosion driven by the second pulse.
In PIC simulations, a set of computational particles is moved under the action of their self-consistent electromagnetic field and any externally applied field: this is done by first depositing the current density on a spatial grid, then solving Maxwell's equations on the same grid and computing the force accelerating each particle, by interpolation of the field values on the position of the point particle.

We first consider the simulation of the irradiation of a cluster having radius $R_0=32$ nm and density $n_0=4.56 \times 10^{22} \text{ cm}^{-3}$ with a pulse sequence whose parameters are the optimal ones calculated in Section \ref{sec:intracluster} ($I_{1} = 8.6 \times 10^{16}$ $\text{W}/\text{cm}^2$, $\Delta t = 236$ fs), except for the peak intensity of the first laser, which has been lowered to $I_1 \simeq 2\times 10^{16}$ $\text{W}/\text{cm}^2$, to compensate for the underestimation of the expansion velocity in the 1D model. The second pulse, having peak intensity $I_2 \simeq 1.3 \times 10^{19}$ $\text{W}/\text{cm}^2$ and a shorter pulse duration ($t\ped{rise}=20$ fs), hits the cluster with time delay $\Delta t=236$ fs (same as the optimal value calculated above). Single-pulse 3D simulations had already shown that peak intensities lower than $I_2$ suffice to expel all cluster electrons before the peak of the pulse reaches the cluster \cite{Peano}, even though the electron dynamics cannot be assumed as instantaneous on the explosion time scale. Both pulses are linearly polarized along the $\hat{z}$ direction, propagate in the $\hat{x}$ direction, and their envelope is approximately Gaussian. The simulation box is cubic, with side $L\ped{box}=1$ $\mu$m, discretized in a $420\times420\times420$ uniform spatial grid and the number of particles per species is $6.4\times10^6$, a value close to the actual number of atoms, for the configuration described.

\begin{figure}[ht]
\epsfig{file=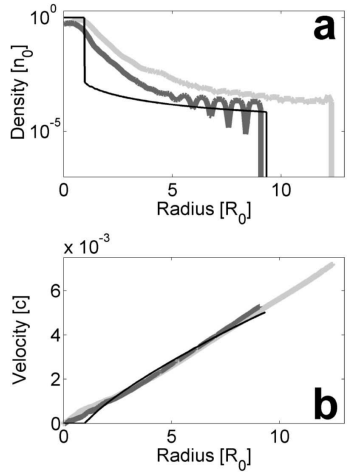, width=7cm}
\caption{(a) Ion number density, and (b) phase space profile at times $t = \Delta t = 236$ fs. In (a), grey thick lines represent lineouts in the $\hat{y}$ (dark) and $\hat{z}$ (light) directions. In (b), grey points mark the position in the $v-r$ phase space for those particles contained in a solid angle $\Delta \Omega \simeq 0.1$ sr around the $\hat{y}$ (dark) and $\hat{z}$ (light) directions. Thin black lines always refer to the solution obtained from the 1D theoretical model.}
\label{fig:density_phase_dt}
\end{figure}

\begin{figure}[ht]
\epsfig{file=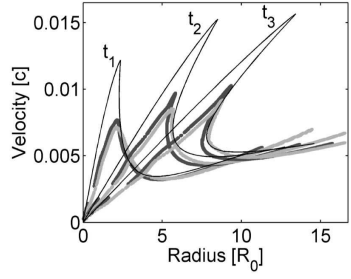, width=7cm}
\caption{Phase space profile at times $t_1 = 270$ fs, $t_2 = 315$ fs, $t_3 = 350$ fs. The gray points mark the position in the $v-r$ phase space for those particles contained in a solid angle $\Delta \Omega \simeq 0.1$ sr around the $\hat{y}$ (dark) and $\hat{z}$ (light) directions. The black lines refer to the solution obtained from the 1D theoretical model.}
\label{fig:phase_all}
\end{figure}

We concentrate our analysis on the ion dynamics in the second part of the simulation, when the more intense laser interacts with the slowly-expanding cluster, driving its Coulomb explosion: right before the interaction with the second pulse begins, the ion density and phase-space profiles appear as in Fig. \ref{fig:density_phase_dt}, where the density profile is decreasing from the center toward the periphery, but the expansion is clearly asymmetric, being much faster along the laser polarization direction, as also testified by the phase space lineouts. As a comparison, the density and velocity profiles obtained with the 1D model are also plotted (in the optimal case $I_{1} = 8.6 \times 10^{16}$ $\text{W}/\text{cm}^2$, $\Delta t = 236$ fs). In the PIC simulation, the cluster has expanded slightly more, despite the lower value of $I_1$ (meaning hydrodynamics effects are indeed very relevant for the configuration considered), and the density profile is different from the theoretical one, being the cluster core less dense. These differences, which are due the limitations of the 1D model, where a crude approximation on the electron dynamics is adopted, do not prevent the formation of a large-scale shock shell. Yet, they appear to affect the Coulomb explosion dynamics, especially in limiting the maximum energy acquired by the inner ions (120 keV instead of the 210 keV predicted by the 1D theory). This is clearly visible in Fig. \ref{fig:phase_all}, where the ion phase-space history is shown. Again, two curves along the $\hat{y}$ and $\hat{z}$ directions are plotted and compared with the theoretical curves. The core explosion is much more symmetric than the first slow expansion, since the electrons are quite rapidly expelled from the cluster and their dynamics has a smaller influence on the ion dynamics than during the first expansion. The explosion predicted by the 1D theory is more violent for a variety of reasons: first, when the second pulse hits the plasma and the explosion starts, the Coulomb energy stored in the cluster core, where much more charge is packed (cf. Fig. \ref{fig:density_phase_dt}), is higher (which also explains why the dynamics of the outer ions, the ones forming the lower branch of the phase space profile, resembles the numerical results more than the dynamics of the inner ions). Furthermore, in the PIC simulations, neutralization by the hot electrons expelled from the cluster, but remaining in the computational domain (where the total net charge is zero), also play a role, as well as 3D effects, anisotropies, boundary effects arising when the ion cloud gets as big as the simulation box, and possible propagation effects in the underdense expanding plasma surrounding the dense cluster core.

The simulations results show that the optimal double-pump configuration predicted by the 1D theory represents a good approximation for inducing the formation of large-shock shells, capable of driving intracluster nuclear reactions in realistic cases, even though all the effects mentioned above can highly affect the expansion/explosion dynamics. Their influence on the reaction yield needs investigation, for instance through numerical calculation of the reaction rates during the PIC simulations, to be presented in a future publication.
However, a first estimate based on Eq. \eqref{eq:rrate2} suggests that differences in the ion dynamics like those of Fig. \ref{fig:phase_all}, where PIC calculations predict a shock shell with velocity amplitude reduced by a factor of two with respect to the 1D model, should result in intracluster reaction yields reduced by a factor of the same order (the actual reduction depends on the variations of $\sigma$ for the conditions considered).
In the case analyzed here, results also seem to suggest that a more pronounced shock shell, with dynamics more akin to the theoretical model, would be obtained with a weaker first pulse and/or a shorter time delay, and with a stronger second pulse. These trends should hold with larger clusters ($R_0 \sim 100-200$ nm) too, provided that conditions lie in the range of validity of the 1D theory.

\section{Conclusions}

The explosion dynamics of large deuterium clusters irradiated by sequential laser pulses having different intensities has been investigated, focusing on the formation and evolution of large-scale shock shells in the ion phase space, which lead to the occurrence of intracluster fusion reactions. The effect of the double-pump parameters (delay of the second pulse and intensity of the first pulse) on the number of intracluster fusion reactions has been analyzed using a 1D model which approximately accounts for the outer ionization dynamics. After the optimal double-pump configuration has been found, for different cluster sizes, the optimal intracluster fusion yield has been calculated and compared with the intercluster yield, finding that intracluster reactions become important with very large clusters (radius $\sim 100$ nm). The optimal double-pump parameters obtained with the 1D model have then been used to perform three dimensional PIC simulations, whose results confirmed the formation of well-pronounced shock shells with high relative velocities inside the single exploding clusters.

\begin{acknowledgments}
The authors would like to thank Prof. G. Coppa, Prof. W. B. Mori, and Dr. F. Peinetti for discussions, and M. Marti and S. Martins for help with the OSIRIS simulations. This work was partially supported by FCT (Portugal) through Grants. PDCT/FP/FAT/50190/2003, and POCI/FIS/55905/2004.
\end{acknowledgments}

\newpage

\end{document}